\documentclass[aps,preprint,11pt]{revtex4} 
\newcommand{\Pc}{{\cal P}}
\newcommand{\Pt}{\tilde{\cal P}}
\newcommand{\tr}{{\rm Tr}}
\newcommand{\rop}{{\hat\rho}}
\newcommand{\xio}{{\hat\xi}}
\newcommand{\eto}{{\hat\eta}}
\begin{document}
%
%\title*{Probability of intrinsic time-arrow\protect\newline  
%        from information loss}  %springer
\title{Probability of intrinsic time-arrow 
        from information loss}
%
%
%\toctitle{Probability of intrinsic time-arrow \protect\newline 
%          from information loss}          %springer
% allows explicit linebreak for the table of content
%
%
%\titlerunning{Probability of intrinsic time-arrow}  %springer
% allows abbreviation of title, if the full title is too long
% to fit in the running head
%
\author{Lajos Di\'osi}
%
%\authorrunning{Lajos Di\'osi}     %springer
% if there are more than two authors,
% please abbreviate author list for running head
%
%
%\institute{Research Institute for Particle and Nuclear Physics\\
%           H-1525 Budapest 114, P.O.Box 49, Hungary}   %springer
\email{diosi@rmki.kfki.hu}\homepage{www.rmki.kfki.hu/~diosi}
\affiliation{Research Institute for Particle and Nuclear Physics\\
           H-1525 Budapest 114, P.O.Box 49, Hungary}

%\maketitle              % typesets the title of the contribution %springer

\begin{abstract}
Time-arrow $s=\pm$, intrinsic to a concrete physical system, is 
associated with the direction of information loss $\Delta I$
displayed by the random evolution of the given system. When the information 
loss tends to zero the intrinsic time-arrow becomes uncertain. We propose
the heuristic relationship $1/[1+exp(-s\Delta I)]$ for the probability
of the intrinsic time-arrow. The main parts of the present work are
trying to confirm this heuristic equation.

The probability of intrinsic time arrow is defined by Bayesian inference 
from the observed random process. From irreversible thermodynamic systems, 
the proposed heuristic probabilities follow via the Gallavotti-Cohen 
relations between time-reversed random processes. In order to explore the 
underlying microscopic mechanism, a trivial microscopic process is analyzed 
and an obvious discrepancy is identified. It can be resolved by quantum
theory. The corresponding trivial quantum process will exactly confirm
the proposed heuristic time-arrow probability.
\end{abstract}

\maketitle

\section{Introduction}
Both experiment and theory confirm that physical processes are time-reversal 
invariant in `simple' systems. This invariance may eventually be lost 
if the system is chaotic, singular, of many degrees of freedom, 
or not isolated \cite{Zeh89}. It seems plausible now that time-reversal
asymmetry (irreversibility) is always accompanied by some
information loss. Yet, little is known quantitatively.   
The present work discusses an elementary informatic mechanism of 
irreversibility. It leads to a simple analytic
expression for the asymmetric probability of the two possible directions of 
time.

Suppose we use \emph{reference-time} $t$ to label the order of events but we 
leave open whether \emph{physical-time} $st$ is passing with increasing 
or decreasing 
$t$, according to the respective \emph{time-arrow} $s=\pm$.
We make no apriori (extrinsic) assignment for $s$.
The ambiguity is to be resolved by analyzing irreversible physical 
processes. We consider informatic irreversibility in a sense that
the Shannon information changes by $\Delta I$ along the process.
We call the resulting aposteriori time-arrow \emph{intrinsic}. 
It belongs to the 
given irreversible process. It would not exist in `empty space' at all. 
In the spirit of the second law of thermodynamics, the physical entropy
production $s\Delta I$ must be positive, hence the intrinsic time-arrow 
is unique: 
\begin{equation}\label{thermo}
s={\rm sign}(\Delta I)~.
\end{equation}
This assignment is only valid if the magnitude $\vert\Delta I\vert$ is
macroscopic which means that it is much bigger than $1$ bit.
If, however, the irreversibility is weak then we have to be contented with 
a probabilistic intrinsic time-arrow. The main suggestion of our work is
that this probabilistic time-arrow is a relevant concept and, furthermore,
the probability $P(s)$ depends on the Shannon information change $\Delta I$
at largely general conditions. I will consider the following relationship:
\begin{equation}\label{heur}
P(s)=\frac{1}{1+e^{-s\Delta I}}~.
\end{equation}
If we change the sign of the reference time $(t\rightarrow-t)$ 
then also the sign of the information loss will change 
$(\Delta I\rightarrow-\Delta I)$. Hence the above expression is
\emph{covariant} against time-reversal of the  reference frame. 
Asymptotically it yields the unique thermodynamic arrow (\ref{thermo}) if 
the information loss $\vert\Delta I\vert$ is much greater than $1$ bit. 
On the contrary, the two time-arrows become equally probable for a 
reversible process where $\Delta I$ is much less than $1$ bit \cite{bit}.      
The suggested relationship is heuristic and lacks a general proof. 
It is intimately related to the fluctuation theorem
\cite{Evaetal93,GalCoh95} 
proved for a particular class of irreversible processes
\cite{Maeetal00}.
On the other hand, it intends to reflect a fundamental
meaning of the time-arrow in terms of information flow.      
I am going to prove that the relationship 
(\ref{heur}) follow from elementary statistical considerations
provided we assume some further conditions to fulfill.

Sec.~\ref{Bayes} presents the mathematical steps of Bayesian 
statistical inference adapted to the estimation of the time-arrow 
from the observed data. In Sec.~\ref{Thermo} we discuss the
inference from irreversible thermodynamic process, 
in Sec.~\ref{Micro} from microscopic process. The time-arrow
is derived from quantum irreversibility in Sec.~\ref{Quant}.
The Appendix offers a short proof of the fluctuation theorem. 

\section{Bayesian time-arrow}\label{Bayes}
Given a statistical system, let $X$ denote a certain random process in a 
given interval of reference-time $t$. 
Let $\tilde X$ denote the time-reversal of $X$. 
Assume that from the principles of statistical physics we  
can calculate the probability $\Pc(X)$ in physical-time! 
We also introduce the probability distribution $\Pt(X)$ of the same random
process seen from a reference frame with reversed time:
\begin{equation}
\Pt(X)\equiv\Pc(\tilde X)~.
\end{equation}
The conditional probability distribution of $X$ takes this form:
\begin{equation}\label{PXs}
P(X\vert s)=
\left\{\begin{array}{ll}\Pc(X)~~~~~&s=+\\
                        \Pt(X)     &s=-\end{array}~~~,\right.
\end{equation}
where $s$ is the apriori time-arrow. Prior to the irreversible process $X$, 
the distribution of $s$ is symmetric: $P_0(s)=1/2$. Hence the joint
distribution of $X$ and $s$ is the following:
\begin{equation}\label{PXsjoint}
P(X,s)=P(X\vert s)P_0(s)=\frac{1}{2}
\left\{\begin{array}{ll}\Pc(X)~~~~~~&s=+\\
                        \Pt(X)     &s=-\end{array}~~~.\right.
\end{equation}
According to the Bayes rule, the conditional 
aposteriori distribution of the time-arrow reads: 
\begin{equation}\label{PsX}
P(s\vert X)=
\frac{1}{\Pc(X)+\Pt(X)}\times
\left\{\begin{array}{ll}\Pc(X)~~~~~&s=+\\
                        \Pt(X)     &s=-\end{array}~~~,\right.
\end{equation}
which can be cast into the following covariant form:
\begin{equation}\label{PsXBay}
P(s\vert X)=
\frac{1}{1+e^{-sD(X)}}~,
\end{equation}
where:
\begin{equation}\label{DX}
D(X)=-\log\frac{\Pt(X)}{\Pc(X)}~.
\end{equation}
This Bayesian estimate means that if 1) we know the apriori distribution 
$\Pc(X)$ of the random process $X$ in physical-time but 2) experimentally 
we observe either $X$ or $\tilde X$ with equal probability since we have no 
apriori information regarding the relationship of our reference-time 
to the physical-time then 3) learning $X$ in the reference-time will 
lead us to the Bayesian probabilistic estimate $P(s)$ of the times-arrow.   

Let us calculate the mean fidelity of the estimated time-arrow: 
from Eqs.~(\ref{PXs}-\ref{DX}) we shall obtain the following closed form:
\begin{equation}\label{F}
F\equiv\sum_X P(+\vert X) P(X\vert+)
=\left\langle\frac{1}{1+e^{-D(X)}}\right\rangle_\Pc~.
\end{equation} 
The expectation value should refer to $\Pc(X)$ which is the distribution
in the physical frame. We can easily derive an ultimate covariant 
expression of the average Bayesian estimate:
\begin{equation}\label{PsBay}
P(s)=
\left\langle\frac{1}{1+e^{-sD(X)}}\right\rangle~,
\end{equation}
where the average refers already to the observed statistics and the
form is valid in time-reversed reference frames as well.

\section{Thermodynamic case}\label{Thermo}
Let $X$ be a coarse-grained macroscopic random process in a
given statistical system in the period $[-T,+T]\equiv[t_1,t_2]$ 
and let $\tilde X$ be the same process seen from the time-reversed
reference frame:
\begin{eqnarray}
X       &=\{X(t); &t_1\leq t\leq t_2\}~,\nonumber\\
\tilde X&=\{X(-t);&t_1\leq t\leq t_2\}~.
\end{eqnarray}
Typically, $X$ can be an irreversible thermodynamic process $X(t)$. 
Assume that we know the irreversible entropy 
$\Delta I(X)$ produced by the process $X$. Obviously, the time-reversed 
process `produces' the same entropy with the opposite sign: 
\begin{equation}
\Delta I(\tilde X)=-\Delta I(X)~.
\end{equation}
Let us introduce the following conditional distributions:
\begin{eqnarray}\label{Pcond}
P(X\vert\xi)        &=&\Pc(X)/P_1(\xi)~,\nonumber\\
P(\tilde X\vert\eta)&=&\Pt(X)/P_2(\eta)~,
\end{eqnarray}
where $P_1(\xi),P_2(\eta)$ are the probability distributions of the extreme
values $\xi=X(t_1)$ and $\eta=X(t_2)$, respectively.
In the Appendix the reader finds an elementary proof of the fluctuation
theorem \cite{Evaetal93,GalCoh95,Maeetal00} 
encoding the asymmetry of the time-reversal $X\leftrightarrow\tilde X$
into covariant equation:
\begin{equation}\label{GC}
P(\tilde X\vert\eta)=e^{-\Delta I(X)}P(X\vert\xi)~.
\end{equation}
Accordingly, the violation of the time-reversal symmetry is exponentially
increasing with the magnitude $\vert\Delta I\vert$ of the irreversible
entropy. We are going to show that, via the Bayesian statistics of
Sec.~\ref{Bayes}, the relationship (\ref{GC}) reproduces the heuristic
probabilities (\ref{heur}) for the thermodynamic time-arrow.

Let us express the r.h.s. of Eq.~(\ref{DX}) from 
Eqs.~(\ref{Pcond},\ref{GC}):
\begin{equation}
D(X)=\Delta I(X)-\log\frac{P_2(\eta)}{P_1(\xi)}~.
\end{equation}
For long enough periods, the r.h.s. is dominated
by the information loss $\Delta I(X)$, the second (boundary) term can be 
ignored (c.f.:\cite{Maeetal00}). In this limit we can write the covariant
Bayesian estimate (\ref{PsXBay}) into this form:
\begin{equation}\label{PsXBayGC}
P(s\vert X)=
\frac{1}{1+e^{-s\Delta I(X)}}~,
\end{equation}
which on average leads to the covariant distribution
\begin{equation}
P(s)=\left\langle\frac{1}{1+e^{-s\Delta I(X)}}\right\rangle~.
\end{equation}
Finally, this yields the heuristic form (\ref{heur}) provided
we can ignore the statistical
fluctuations of the entropy production around its expectation value
$\Delta I=\langle I(X)\rangle$. This is justified for common
macroscopically irreversible processes where $\vert\Delta I\vert\gg1$.

\section{Microscopic case}\label{Micro}
Let us consider a statistical ensemble of $n\gg1$ independent 
$d-$state systems characterized by the probability distribution
$\rho^i, i=1,2,\dots, d$. 
Let $X$ be an abstract random process as trivial as 
the transition from an initial microscopic ensemble state 
$\xi$ into a final one $\eta$, the time-reversed process 
$\tilde X$ will be the opposite transition:
\begin{eqnarray}
X       =(\xi,\eta)~,\nonumber\\
\tilde X=(\eta,\xi)~.
\end{eqnarray}
Let $\rho_1^i$ and $\rho_2^i$ be the systems' probability 
distributions within the ensembles $\xi$ and $\eta$, respectively.
Then the change of Shannon information along the process $X$ reads: 
\begin{equation}\label{DI}
\Delta I\equiv nI_2-nI_1=-n\sum_{i=1}^d \rho_2^i \log\rho_2^i
                         +n\sum_{i=1}^d \rho_1^i \log\rho_1^i~.
\end{equation}
The process $X$ is irreversible if $\Delta I\neq0$ and we should assign 
the time-arrow $s$ so that $s\Delta I$ be positive (\ref{thermo}). 
The point is that the two samples $\xi$ and $\eta$ may, by
chance, not realize the asymmetry especially when the shapes of their
probability distributions $\rho_1^i$ and $\rho_2^i$ do not much differ
from each other.

Let us characterize the
two constituting configurations of $X=(\xi,\eta)$ by the multiplicities
$n_1^i$ and $n_2^i$:
\begin{eqnarray}
\xi =(n_1^i;i=1,2,\dots,d)~,\nonumber\\
\eta=(n_2^i;i=1,2,\dots,d)~,
\end{eqnarray}
which follow independent multinomial distributions with the
respective mean values
\begin{eqnarray}
\langle n_1^i\rangle=n\rho_1^i~,\nonumber\\
\langle n_2^i\rangle=n\rho_2^i~.
\end{eqnarray}
For large $n$ we can approximate the multinomial distributions
by Gaussian functions:
\begin{eqnarray}
\Pc(\xi,\eta)=C\exp
\left(-\sum_{i=1}^d \frac{[n_1^i-n\rho_1^i]^2}{2n\rho_1^i}
      -\sum_{i=1}^d \frac{[n_2^i-n\rho_2^i]^2}{2n\rho_2^i}
\right)~,\nonumber\\
\Pc(\eta,\xi)=C\exp
\left(-\sum_{i=1}^d \frac{[n_2^i-n\rho_1^i]^2}{2n\rho_1^i}
      -\sum_{i=1}^d \frac{[n_1^i-n\rho_2^i]^2}{2n\rho_2^i}
\right)~.
\end{eqnarray}
We substitute these expressions into Eq.~(\ref{DX}) to calculate
$D(\xi,\eta)$, then we calculate the mean value:
\begin{equation}\label{D}
D=-\frac{n}{2}\sum_{i=1}^d\left((\rho_2^i)^2-(\rho_1^i)^2\right)
     \left(\frac{1}{\rho_2^i}-\frac{1}{\rho_1^i}\right)~.
\end{equation}
Suppose that $D(\xi,\eta)$ is, for very large $n$, 
dominated by the mean value $D$ and fluctuations
will thus be ignored. Hence the average Bayes estimate (\ref{PsBay}) reads:
\begin{equation}
P(s)=\frac{1}{1+e^{-sD}}~. 
\end{equation}
This could become equivalent with our heuristic proposal provided
$D=\Delta I$ which is apparently not true in general.
I was looking for further conditions at least to achieve
the asymptotic equivalence of $D$ and $\Delta I$.
I concluded to the following elementary assumptions. First,
the shapes $\rho_1^i$ and $\rho_2^i$ must be close to
each other so that the lowest nontrivial order in 
$\Delta\rho^i=\rho_2^i-\rho_1^i$ will be sufficient.
Second, the statistics of \emph{either} $\xi$ \emph{or} $\eta$ must be 
totally random. This sets the apriori time-arrow for $s=-$ or $s=+$,
respectively. For concreteness, I consider the case $s=+$ and adopt flat
distribution for $\eta$ \cite{cov}: 
\begin{equation}\label{flat}
\rho_2^i=\frac{1}{d}~.
\end{equation}
This second assumption is a necessary one, otherwise $\Delta I$ 
contains a linear term in $\Delta\rho^i$ while $D$ does not.
From Eqs.~(\ref{DI}) and (\ref{D}) the above two assumptions lead to 
the following results:
\begin{equation}\label{DIapprox}
\Delta I=\frac{nd}{2}\sum_{i=1}^d (\Delta\rho^i)^2~,
\end{equation}
and
\begin{equation}\label{Dapprox}
D =nd\sum_{i=1}^d (\Delta\rho^i)^2~.
\end{equation}
The result is surprising: $D$ has come out twice the information loss. 

Mathematically, $D$ is the Kullback divergence between two
neighboring ensembles $\xi$ and  $\eta$ and it should 
asymptotically coincide with the information loss between them. 
The reason of the anomalous factor $2$ is that we happened to use 
the Kullback divergence between
the composite ensembles $(\xi,\eta)$ and $(\eta,\xi)$ instead of
$\xi$ and $\eta$. This gives a hint how the factor $2$ would go
away. It is interesting to note that the physical resolution has
a typical quantum mechanical motivation. In microphysics it is
conceptually impossible to observe the full quantity $X=(\xi,\eta)$.
If, e.g., the time-arrow is positive $(s=+)$ then $\eta$ is
testable and $\xi$ is not because its observation would significantly
perturb the initial preparation. And vice versa, when $s=-$ then 
$\eta$ is testable and $\xi$ is not. Accordingly, we are going
to change the concept of experimental data. In the concrete case, 
we forbid the observation of $\xi$. In this sense, we have to 
redefine the distribution of the observed quantities:
\begin{eqnarray}
\Pc(\xi,\eta)\rightarrow\Pc(\eta)\equiv\sum_\xi\Pc(\xi,\eta)~,
\nonumber\\
\Pt(\xi,\eta)\rightarrow\Pt(\eta)\equiv\sum_\xi\Pt(\xi,\eta)~.
\end{eqnarray}
Repeating the calculation of the Kullback divergence in the leading
order, inserting the flat values (\ref{flat}) for $\rho_2^i$, 
$D$ turns out to be half of the previous value (\ref{DIapprox}).
Thus in the given approximation we have obtained the identity
\begin{equation}
D=\Delta I~,
\end{equation}
and confirmed the heuristic relationship (\ref{heur}).

\section{Quantum case}\label{Quant}
Let us consider the statistical ensemble of $n\gg1$ independent 
$d-$state quantum systems where each one has the same density
matrix $\rop$. 
Let $X$ be an abstract random process as trivial as 
the transition from an initial ensemble $\xio$ into
a final one $\eto$, the time-reversed process $\tilde X$ will be 
the opposite transition: 
\begin{eqnarray}
X       =(\xio,\eto)~,\nonumber\\
\tilde X=(\eto,\xio)~,
\end{eqnarray}
where
\begin{eqnarray}\label{xieta}
\xio &=\rop_1\otimes\rop_1\otimes\dots\otimes\rop_1
                                             \equiv\rop_1^{\otimes n}~,
\nonumber\\
\eto&=\rop_2\otimes\rop_2\otimes\dots\otimes\rop_2
                                             \equiv\rop_2^{\otimes n}~,
\end{eqnarray}
if $\rop_1$ and $\rop_2$ stand for the systems' density matrices 
within the ensembles $\xio$ and $\eto$, respectively.
The change of von Neumann information during the process $X$ reads: 
\begin{equation}\label{DIvN}
\Delta I\equiv nI_2-nI_1=-n\tr\left(\rop_2 \log\rop_2\right)
                         +n\tr\left(\rop_1 \log\rop_1\right)~.
\end{equation}
The process $X$ is irreversible if $\Delta I\neq0$ and we should assign 
the time-arrow $s$ so that $s\Delta I$ be positive (\ref{thermo}). 
In order to $\Delta I$ have a definite sign
the two ensembles $\xio$ and $\eto$ should display experimentally 
significant asymmetry.
 
Quantum theory says that if the reference-time is the physical
time $(s=+)$ then we can not test the ensemble $\xio$ but
the ensemble $\eto$. And in the opposite case $(s=-)$ the
ensemble $\xio$ is testable and $\eto$ is not. We see that the estimation
of the time-arrow $s$ boils down to the statistical decision whether
the actually observed ensemble is $\xio=\rop_1^{\otimes n}$ 
or $\eto=\rop_2^{\otimes n}$ 
whereas both alternatives have equal apriori likelihoods.   

We can mechanically follow the Bayes method of the previous chapters.
Note, however, the typical quantum informatic arguments: this is the way 
I approached the issue originally.

The two collective states (\ref{xieta}) reside in a Hilbert space of 
dimension $d^n$. According to the quantum counterpart of Shannon's code 
theory \cite{Sch95}, in the large $n$ limit such collective states 
become asymptotically equivalent with totally random states 
restricted for given subspaces. Our states (\ref{xieta}) become random
states in subspaces $\hat E_1$ and $\hat E_2$:
\begin{eqnarray}\label{xietaasymp}
\xio&=&\rop_1^{\otimes n}\sim e^{-nI_1}\hat E_1~,\nonumber\\
\eto&=&\rop_2^{\otimes n}\sim e^{-nI_2}\hat E_2~,
\end{eqnarray}
where $\hat E_1$ and $\hat E_2$ are Hermitian projectors of dimensions
$e^{nI_1}$ and $e^{nI_2}$, respectively. The dimensions depend
on the von Neumann entropies. We are interested in the situations
where the experimental distinguishability of the above two ensembles
would exclusively depend on the difference $\Delta I=nI_2-nI_1$ of the
informations. This is obviously not true in general because the
distinguishability will depend e.g. on the overlap 
$\tr(\hat E_1 \hat E_2)$.
Nonetheless, the asymptotic forms (\ref{xietaasymp}) suggest simple
conditions to achieve our goal. Suppose that one of the ensembles,
say $\eto$, is of maximum information:
\begin{equation}
\rop_2=\frac{\hat 1}{d}~,
\end{equation}
which means that $I_2=\log d$ and $\hat E_2=\hat 1^{\otimes n}$. 
The ensemble 
$\eto$ is totally random over the whole collective Hilbert space of
dimension $d^n$. The overlap between $\hat E_1$ and $\hat E_2$ 
becomes trivial.
The information loss is always positive and the true time-arrow is
thus $s=+$. But we have to find it by deciding whether we have tested
the ensemble $\xio$ or $\eto$ which are of equal apriori likelihoods.

Now the experimental distinguishability of $\xio$ and $\eto$ is already 
trivial. All we have to do is to define $\hat E_1$ as observable and
to observe it! If the tested ensemble is $\xio$ itself then we get
$1$ with certainty since $\tr(\hat E_1\xio)=1$. If the observed ensemble is 
the fully random $\eto$ then we get $1$ with probability 
$\tr(\hat E_1\eto)=e^{-\Delta I}$ and we get $0$ with the
complementer probability. As we see, the complete experimental statistics
is determined by the information loss $\Delta I$.

Let us turn to the Bayes method of Sec.~\ref{Bayes} to estimate the 
time-arrow $s$.
As we suggested above, the observed data is the value $E_1=\{0,1\}$ of
the quantum observable $\hat E_1$. The probability $\Pc(E_1)$ stands for 
its distribution in the 
reference time with time-arrow $s=+$ and $\Pt(E_1)$ stands for its
distribution in the reversed time $s=-$. In the preceding paragraph
we established their values:   
\begin{eqnarray}\label{PE_1}
\Pc(E_1)&=&E_1e^{-\Delta I}+(1-E_1)(1-e^{-\Delta I})~,\nonumber\\
\Pt(E_1)&=&E_1~.
\end{eqnarray}
Applying the steps of Sec.~\ref{Bayes} mechanically, first we write 
Eq.~(\ref{DX}) into this form:
\begin{equation}
e^{-D(E_1)}=\frac{\Pt(E_1)}{\Pc(E_1)}~,
\end{equation}
which is then substituted into the expression (\ref{F}) of the mean
fidelity, yielding:
\begin{equation}
F=\left\langle\frac{1}{1+\Pt(E_1)/\Pc(E_1)}\right\rangle_\Pc
            =\frac{1}{1+e^{-\Delta I}}
\end{equation} 
We have used Eqs.~(\ref{PE_1}) to calculate the average. 
The result implies exactly the form (\ref{heur}) for the probability
of intrinsic time-arrow in function of the information loss.

\section{Concluding remarks}
I proposed a heuristic probability distribution (\ref{heur}) for the
time-arrow intrinsic to a given irreversible process. The proposed
probability is solely a function of the information loss $\Delta I$.
The idea itself comes from the the phenomenological fluctuation theorem.
Indeed, the concrete form of my proposal can easily be confirmed for
the intrinsic time-arrow of standard irreversible processes, at least
in the limit of macroscopic entropy production $\Delta I\gg1$. My basic
goal, however, was the construction of whatever trivial microscopic 
process which could underly the proposed dependence on $\Delta I$.
I analyses the irreversible process of the simplest possible
structure in classical and quantum versions. The quantum version
confirmed the proposed probabilistic time-arrow. Let me summarize
this central result. 

1) Suppose we know that (in physical time) a quantum ensemble $\xio$ of
$n\gg1$ identical systems of given (also known to us) state transforms 
into an ensemble $\eto$ of $n$ totally random systems. 
2) Suppose we do not know at all whether our reference-time is the 
physical-time 
or not, and whether the `resulting ensemble' of the above process has
been $\eto$ or $\xio$. 3) We test the `resulting ensemble' and Bayesian
inference will give us the time-arrow with fidelity
$$
F=\frac{1}{1+e^{-\vert\Delta I\vert}}~.
$$

An infite number of conceptual issues could be raised against the 
presented ideas. I mention and discuss only two. First, the
assignment of a non-trivial intrinsic time-arrow to a local irreversible
process is a speculation. Nature might retain the same universal 
time-arrow for the whole Universe independently of the 
measure or direction of local information flows. Yet, we do not know
if Nature is that conservative indeed. We learned from Einstein that Nature
delegates the issues of local geometry to local physical systems.
I adopted the hypothesis that this happens with time-arrow as well.
Second, the proposed confirmation of the time-arrow probability includes
Bayesian inference. Many would say that inference is subjective. The
obtained probability is also subjective. Nonetheless, famous arguments 
using inference have been used earlier to confirm objective statistics of
quantized fields \cite{BohRos33}. It is, furthermore, a common knowledge 
that the maximum-likelihood inference of the intensive thermodynamic
parameters confirms their true equilibrium fluctuations in Gibbs
ensembles. 

The present work is an attempt to find universal expressions for the
hypothetic intrinsic time-arrow. There is a hint of the
information loss to play the key role. This does not mean that we can
already claim an experimental significance which should, of course,
be inevitable after all. But theory of intrinsic time opens
a series of natural questions to study in the future and there is
apparently a promise of further analytic results.

\section*{Acknowledgment(s)}
I am grateful to Hans-Thomas Elze and to the sponsors of the workshop
"DICE 2002" for the invitation to talk in Piombino and
to contribute to this volume. My research was also supported by the
Hungarian OTKA Grant No. 032640.

\appendix
\section*{Appendix}
Let $X(t)$ denote a thermodynamic variable of equilibrium value $\bar X$,
where $\lambda$ is the relaxation rate, and $\gamma$ 
is the Onsager kinetic coefficient.
The time-dependent fluctuations of  
$X(t)$ are governed by the phenomenological Langevin equation:
\begin{equation}\label{Lan}
\frac{dX(t)}{dt} = -\lambda(X(t)-{\bar X}) 
           +\sqrt{2\gamma}~w(t) 
\end{equation}
with the standard white-noise $w(t)$. The expression 
\begin{equation}\label{Irate}
\frac{dI}{dt}=\frac{\lambda}{\gamma}(\bar X-X)\frac{dX}{dt}
\end{equation}
will be the local rate of irreversible entropy production (information loss)
along the process $X(t)$ (see, e.g., in Landau-Lifshitz \cite{LanLif82}, or
in \cite{Dioetal96}).
According to Onsager and Machlup \cite{OnsMac53}, 
the conditional probability distribution of the process 
\mbox{$X=\{X(t);-T\leq t\leq T\}$}
at fixed initial value $\xi=X(-T)$ and for equilibrium value $\bar X$
takes this functional Gaussian form:
\begin{equation}\label{OM}
P(X\vert\xi;\bar X)=\exp\Bigl(
-{1\over4\gamma}\int_{-T}^{T}
\left[\frac{dX(t)}{dt} +\lambda(X(t)-{\bar X})\right]^2dt\Bigr)~.
\end{equation}
We shall consider \emph{driven} thermodynamic processes which can be
described by the Eqs.~(\ref{Lan}-\ref{OM}) with  
time-dependent equilibrium values 
\mbox{$\{{\bar X}(t);-T\leq t\leq T\}$}. For convenience
of forthcoming calculations let us write down the distribution
functional of the driven process:
\begin{equation}\label{OMdriv}
P(X\vert\xi;{\bar X})=\exp\Bigl(-{1\over4\gamma}\int_{-T}^{T}
\left[\frac{dX(t)}{dt} +\lambda(X(t)-{\bar X}(t))\right]^2dt\Bigr)~.
\end{equation}
Obviously the above equations assume physical time $t$. Let us express 
the conditional distribution of the time-reversed process $\tilde X$ 
starting from $\tilde X(-T)=\eta$, driven by the time-reversed 
function $\tilde{\bar X}$. Namely, we replace $X,\xi,\bar X$ in 
Eq.~(\ref{OMdriv}) by $\tilde X,\eta,{\tilde{\bar X}}$, respectively:
\begin{equation}\label{OMdrivrev}
P(\tilde X\vert\eta;\tilde{\bar X})
=\exp\Bigl(-{1\over4\gamma}\int_{-T}^{T}
\left[\frac{d\tilde X(t)}{dt} +\lambda(\tilde X(t)-\tilde{\bar X}(t))\right]
^2dt\Bigr)~.
\end{equation}
Now we change the variable $t$ in the integrand for $-t$ and insert the
relations:
\begin{eqnarray}
\tilde       X(t)\equiv     X(-t)~,\nonumber\\
\tilde{\bar X}(t)\equiv\bar X(-t)~,
\end{eqnarray}
leading to:
\begin{equation}\label{OMdrivrev1}
P(\tilde X\vert\eta;\tilde{\bar X})
=\exp\Bigl(-{1\over4\gamma}\int_{-T}^{T}
\left[\frac{dX(t)}{dt} -\lambda(X(t)-\bar X(t))\right]
^2dt\Bigr)~.
\end{equation}
(Recall that this expression would be the conditional distribution of the 
process had we observed it in the time-reversed frame.)
The logarithm of the physical distribution (\ref{OMdriv}) over the 
time-reversed one (\ref{OMdrivrev1})
will result in a remarkable expression:
\begin{equation}
\log\frac{P(X\vert\xi;\bar X)}{P(\tilde X\vert\eta;\tilde{\bar X})}=
\frac{\lambda}{\gamma}\int_{-T}^T (\bar X(t)-X(t))dX(t)~.
\end{equation}
It follows from Eq.~(\ref{Irate}) that the r.h.s. is equal to the
total entropy production (information loss) of the driven process:
\begin{equation}
\Delta I(X;\bar X)
=\frac{\lambda}{\gamma}\int_{-T}^T(\bar X(t)-X(t))dX(t)~.
\end{equation}
This and the preceding equation yield the fluctuation theorem
\cite{Evaetal93}-\cite{Maeetal00}:
\begin{equation}
P(\tilde X\vert\eta;\tilde{\bar X})=e^{-\Delta I(X;\bar X)}
P(       X\vert\xi ;       \bar X )~.
\end{equation}

%INDEX%%%%%%%%%%%%%%%%%%%%%%%%%%%%%%%%%%%%%%%%%%%%%%%%%%%%%%%%%%%%%%%

% Please check with the editor of your book whether he plans to

% include a "mutual" subject index - if so, please code your entries

% in the standard syntax. For your own purposes you may print your

% "personal" index by using the following commands:

%

%\clearpage

%\addcontentsline{toc}{section}{Index}

%\flushbottom

%\printindex

%%%%%%%%%%%%%%%%%%%%%%%%%%%%%%%%%%%%%%%%%%%%%%%%%%%%%%%%%%%%%%%%%%%%%


\begin{thebibliography}{8.}
\bibitem{bit} Especially in thermodynamic context, the Shannon and
von Neumann information (loss) will be deliberately identified with 
the entropy (production). While information is the fundamental quantity 
in the present work, I use Boltzmann units of entropy. In Shannon units 
(bits) the central equation (\ref{heur}) would read:
$$
P(s)=\frac{1}{1+2^{-s\Delta I}}.
$$
\bibitem{cov} The choice only means that, for simplicity's sake,
the forthcoming calculations are done in the physical reference time.
The covariance for $t\rightarrow-t$ will only be restored on the final 
equations.
\end{thebibliography}
\end{document}